# Ultrastrong coupling between magnetoplasmons and higher cyclotron-resonance harmonics induced by spatially inhomogeneous electromagnetic fields in two-dimensional electron systems

Kazuyuki Kuroyama[1], Shimpei Nakamura[2], and Motoaki Bamba[2,3,4]

[1]*Institute of Industrial Science, The University of Tokyo, Tokyo, Japan*

[2]*Department of Physics, College of Engineering Science, Yokohama National University, Kanagawa, Japan*

[3]*Department of Physics, Graduate School of Engineering Science, Yokohama National University, Kanagawa, Japan*

[4]*Institute for Multidisciplinary Sciences, Yokohama National University, Kanagawa, Japan*

*corresponding author: kuroyama@iis.u-tokyo.ac.jp

## Abstract

Finite-wave-vector electromagnetic fields provide a route to accessing nonlocal collective dynamics in two-dimensional electron systems. Here, we theoretically investigate the magnetoplasmon response of a two-dimensional electron system subjected to a spatially nonuniform electromagnetic field in a perpendicular magnetic field. By combining Maxwell's equations with the Kubo formula, we derive the nonlocal magnetoconductivity tensor and obtain the self-consistent dispersion relation for collective modes. We show that a finite in-plane wave vector enables higher-order cyclotron transitions, allowing higher harmonics of the cyclotron resonance to contribute to the nonlocal conductivity. These harmonics hybridize with the magnetoplasmon mode, producing avoided crossings in the collective-mode spectrum characteristic of Bernstein modes. To quantify this hybridization, we establish a correspondence between the Maxwell–Kubo dispersion relation and a Hopfield-type Hamiltonian and derive an analytical expression for the effective coupling strength between the magnetoplasmon and the $n$-th cyclotron-resonance harmonic. Our analysis reveals that the normalized coupling strength exhibits a nonmonotonic dependence on the in-plane wavelength, reaching a maximum that becomes nearly independent of electron density in the high-density regime. These results establish a quantitative connection between nonlocal magnetoconductivity, Bernstein-mode physics, and effective collective-mode coupling, providing design guidelines for realizing ultrastrong coupling in terahertz magnetoplasmonic systems.

# I. Introduction

Collective electrodynamic excitations in two-dimensional electron systems (2DESs) have provided a fundamental platform for studying the interplay between charge dynamics, electromagnetic fields, and magnetic-field-induced cyclotron motion [1,2]. In a translationally invariant 2DES with a parabolic energy dispersion, Kohn's theorem constrains the response to a spatially uniform electric field, such that the cyclotron resonance is unaffected by electron–electron interactions [3]. However, electromagnetic fields with finite in-plane wave vectors can probe internal electronic dynamics beyond the uniform center-of-mass response. Such nonlocal electrodynamic effects provide opportunities to explore collective excitations and their coupling to electromagnetic fields beyond the long-wavelength limit. This nonlocal interaction exposes a rich variety of hidden phenomena, including strong electron–electron correlations and optical nonlinearities, which influence the quantum Hall effect and hold the potential for breakthroughs in novel quantum devices.

In a perpendicular magnetic field, collective plasmon excitations in a two-dimensional electron system couple to cyclotron motion, giving rise to magnetoplasmon modes whose frequencies depend on both the bare plasmon frequency and the cyclotron frequency [4]. As extensively studied in semiconductor heterostructures, this magnetoplasmon response represents a fundamental collective mode emerging from the self-consistent interaction between electron density oscillations and electromagnetic fields. Furthermore, the spatial modulation of the electric field probes the nonlocal conductivity of the 2DES [4,5], enabling transitions between Landau levels separated by multiple cyclotron quanta. Consequently, magnetoplasmons can hybridize with these higher cyclotron-resonance (CR) harmonics, exhibiting avoided crossings in the spectra. This phenomenon is closely related to Bernstein-mode physics in two-dimensional systems [4–14].

The coupling between magnetoplasmons and higher CR harmonics has attracted renewed interest in connection with THz near-field excitation and ultrastrong coupling [15–17]. Recent experiments have demonstrated that this coupling can be enhanced by spatially confining the electromagnetic near field [13]. Furthermore, theoretical studies have predicted the emergence of additional collective excitation modes in strongly interacting 2DESs [18,19]. Despite these advances, experimental evidence of modifications to the many-body dynamics of correlated electrons induced by ultrastrong light–matter coupling remains limited [20]. In addition to recent theoretical efforts to elucidate these mechanisms [21,22], establishing a more comprehensive understanding of these complex dynamics remains essential to drive further advancements in the field [23,24] .

In this work, we theoretically investigate the finite-wave-vector magnetoplasmon response of a 2DES by combining Maxwell's equations with the Kubo formula. We consider an electromagnetic field with a finite in-plane wave vector $q_x$, which can be introduced, for example, by a diffraction grating with period $L$, where the first spatial harmonic corresponds to $q_x = 2\pi/L$ [7,8,11,25]. We first obtain the self-consistent dispersion equation for the collective modes using the nonlocal Kubo conductivity tensor $\sigma_{ij}(q_x, \omega)$ in a perpendicular magnetic field. The resulting dispersion shows avoided crossings between the magnetoplasmon mode and higher harmonics of the cyclotron resonance. We then establish a correspondence between the Maxwell–Kubo dispersion relation and a Hopfield-type coupled-oscillator Hamiltonian describing the interaction between the magnetoplasmon and higher cyclotron harmonics. In the nonretarded and weak-dissipation limits, this correspondence allows us to derive an analytical expression for the effective coupling strength $g_n$ between the magnetoplasmon and the $n$-th cyclotron-resonance harmonic.

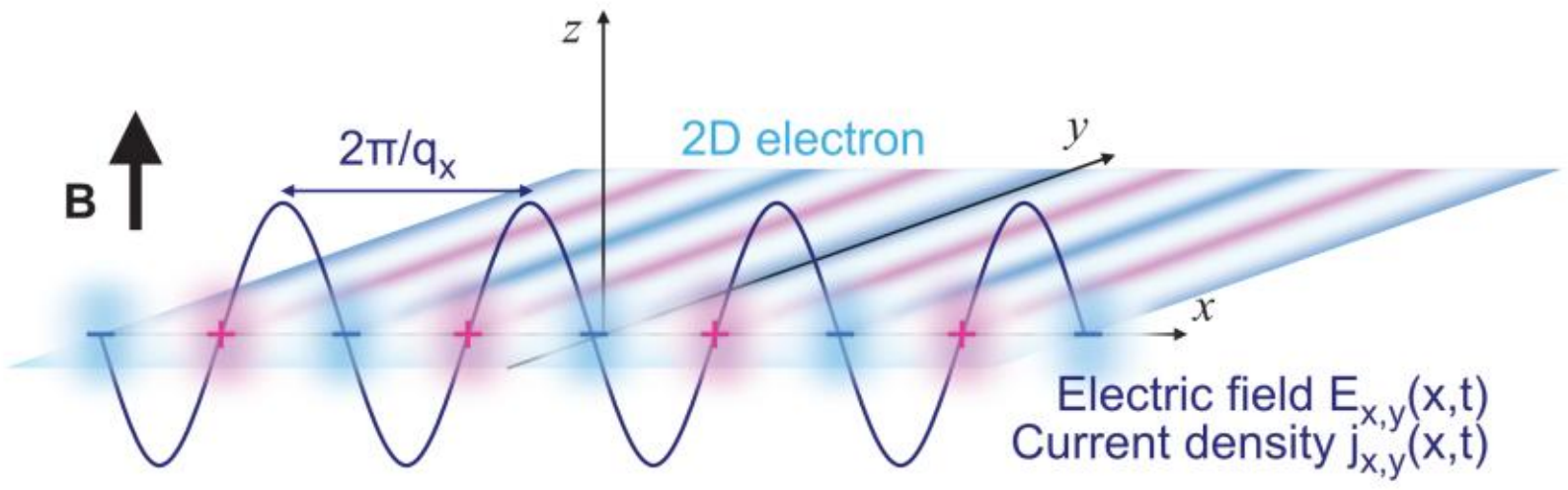

Fig. 1. Geometry of the uniform 2D electron plane at $z = 0$ plane exposed to the diffracted electric field. The magnetic field **B** is applied along the $z$-axis.

Using this analytical expression, we investigate the dependence of the normalized coupling strength on the in-plane wavelength and electron density. We find that the coupling strength exhibits a nonmonotonic dependence on the wavelength, reaching a maximum at a characteristic wavelength that becomes nearly independent of electron density in the high-density regime. The maximum normalized coupling strength can exceed the conventional threshold for ultrastrong coupling. These results establish a quantitative connection between nonlocal magnetoconductivity, Bernstein-mode physics, and effective coupled-oscillator descriptions, and provide design guidelines for enhancing the coupling between magnetoplasmons and higher cyclotron harmonics through engineered finite-wave-vector electromagnetic fields.

# II. Model and nonlocal dispersion relation

We consider a uniform 2DES located in the $xy$ plane at $z = 0$ in a semiconductor, as schematically shown in Fig. 1. The finite thickness of the electron layer is neglected, and the 2DES is treated as a conducting sheet. A magnetic field **B** is applied along the $z$ axis. We investigate the collective eigenmodes of the 2DES at a finite in-plane wave vector $q_x$. Accordingly, we assume translational invariance along the $y$ direction, so that all physical quantities are independent of $y$. In the local long-wavelength limit, the electrodynamic response of a parabolic-band 2DES is governed primarily by the ordinary cyclotron resonance. At finite $q_x$, however, the response becomes nonlocal, and higher harmonics of the cyclotron resonance acquire finite oscillator strength. These higher harmonics can then hybridize with the magnetoplasmon mode, giving rise to avoided crossings in the collective-mode spectrum.

Following Ref. [26], the electric field satisfies the wave equation

$$\nabla \times [\nabla \times \mathbf{E}(\mathbf{r}, t)] + \varepsilon\mu \frac{\partial^2}{\partial t^2} \mathbf{E}(\mathbf{r}, t) = -\mu \frac{\partial}{\partial t} [\mathbf{j}(x, t)\delta(z)]. \tag{1}$$

Here, $\mathbf{j}(x, t)$ is the two-dimensional sheet current density, and $\delta(z)$ localizes the current to the 2DES plane. The surrounding dielectric environment is represented by the effective permittivity $\varepsilon$ and permeability $\mu$. Solving Eq. (1) for a sheet current at $z = 0$ gives the self-consistent relation between the induced electric field and the sheet current density,

$$E_x(q_x, \omega) = -\frac{i\kappa}{2\varepsilon\omega} j_x(q_x, \omega), \tag{2}$$

and

$$E_y(q_x, \omega) = \frac{i\mu\omega}{2\kappa} j_y(q_x, \omega), \tag{3}$$

where $E_x(q_x, \omega)$ and $E_y(q_x, \omega)$ are the Fourier transform of the electric field at $z = 0$, and

$$\kappa(q_x, \omega) \equiv \sqrt{q_x^2 - \varepsilon\mu\omega^2} \tag{4}$$

with Re $\kappa > 0$. The derivations of Eqs. (2) and (3) are given in Appendix A.

We next calculate the nonlocal magnetoconductivity using the Kubo formula. According to the linear response theory, the finite-wave-vector conductivity tensor is obtained as

$$\sigma_{\alpha\beta}(q, \omega) = \frac{i}{\omega}\left[\frac{e^2 n_s}{m^*}\delta_{\alpha\beta} + \chi_{j_\alpha^p j_\beta^p}(\mathbf{q}, \omega)\right], \tag{5}$$

where $m^*$ is the electron's effective mass, $n_s$ is the electron sheet density, and $\chi_{j_\alpha^p j_\beta^p}$ is the paramagnetic current–current response function using the Landau-level basis. The Lehmann representation of the retarded response function for Landau levels is

$$\begin{aligned}\chi_{j_\alpha^p j_\beta^p}(\mathbf{q}, \omega) &= \frac{1}{S}\sum_{n,m}\sum_{k,k'} \frac{f_n - f_m}{\hbar\omega + \varepsilon_n - \varepsilon_m} \langle nk | j_\alpha^p(\mathbf{q}) | mk' \rangle \left\langle mk' \middle| j_\beta^p(-\mathbf{q}) \middle| nk \right\rangle \\ &= \frac{1}{S}\sum_{n,m}\sum_{k,k'} \frac{f_n - f_m}{\hbar\omega + (n-m)\hbar\omega_c} \langle nk | j_\alpha^p(\mathbf{q}) | mk' \rangle \left\langle mk' \middle| j_\beta^p(-\mathbf{q}) \middle| nk \right\rangle, \end{aligned} \tag{6}$$

where $\omega_c = eB/m^*$ is the cyclotron frequency, and $f_n$ is the occupation factor of the $n$-th Landau level. The paramagnetic current-density operator in momentum space is

$$\mathbf{j}^p(\mathbf{q}) = -\frac{e}{2m^*}\left(\mathbf{\Pi} e^{-i\mathbf{q}\cdot\mathbf{r}} + e^{-i\mathbf{q}\cdot\mathbf{r}}\mathbf{\Pi}\right), \tag{7}$$

where the kinetic momentum $\mathbf{\Pi} = (\Pi_x, \Pi_y)$ is defined as

$$\Pi_x = p_x, \tag{8}$$
$$\Pi_y = p_y + eBx. \tag{9}$$

Then, the longitudinal and Hall conductivities are

$$\sigma_{xx}(q_x, \omega) = \frac{i}{\omega}\left[\frac{e^2 n_s}{m^*} + \frac{e^2\omega_c^2}{\pi\hbar l_B^2}\sum_{n,m}\frac{f_n - f_m}{\omega + (n-m)\omega_c}\left(\frac{n-m}{q_x}\right)^2 F_{nm}(q_x)F_{mn}(-q_x)\right], \tag{10}$$

and

$$\sigma_{xy}(q_x, \omega) = -\frac{e^2\omega_c^2}{\pi\hbar\omega l_B^2}\sum_{n,m}\frac{f_n - f_m}{\omega + (n-m)\omega_c}\left(\frac{n-m}{q_x}\right) F_{nm}(q_x)\frac{\partial}{\partial q_x}F_{mn}(-q_x), \tag{11}$$

where $l_B = \sqrt{\hbar/(eB)}$ is the magnetic length. It should be noted that we consider the spin degeneracy but not the Zeeman spin splitting. The expressions of $\sigma_{yx}(q_x, \omega)$ and $\sigma_{yy}(q_x, \omega)$ are also given in Appendix B. The Landau-level form factor $F_{nm}(q_x)$ is defined as

$$F_{nm}(q_x) = \frac{1}{\sqrt{2^{n+m}\pi n!\, m!}}\int du\, e^{-iq_x l_B u - u^2} H_n(u) H_m(u), \tag{12}$$

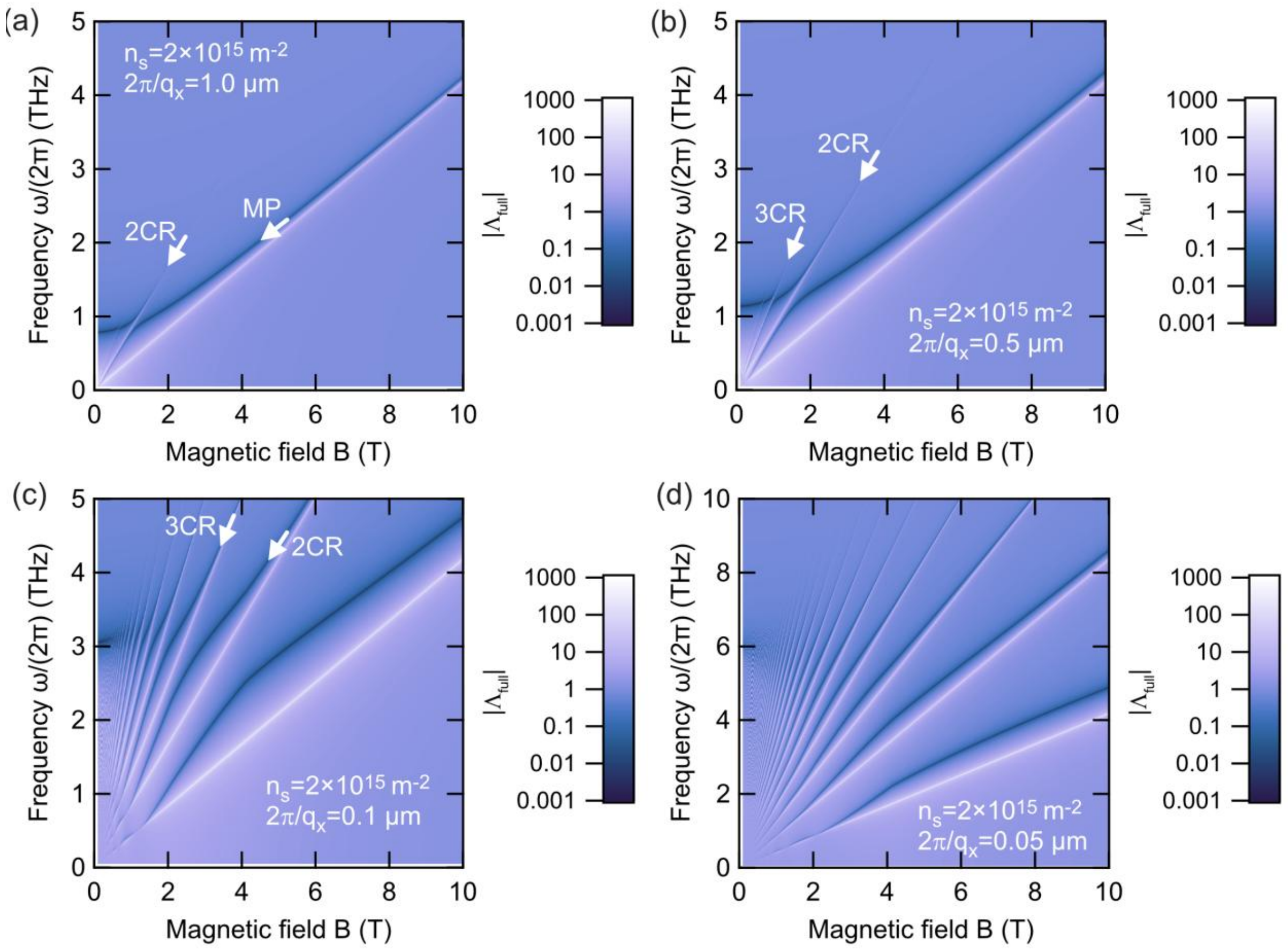

Fig. 2. Color-coded maps of $|\Lambda_{\text{full}}(q_x, \omega)|$ for the different values of $2\pi/q_x$ (the values are written in the plots) at the electron temperature of 0 K. The dark regions in the color maps correspond to the solutions of the dispersion equation in Eq. (16).

where $H_n(u)$ is the Hermite polynomial of the $n$-th order. The derivations of the conductivities are summarized in Appendix B. These expressions explicitly show that the finite-wave-vector conductivity has resonant structures at $\omega = n\omega_c$. Thus, the higher harmonics of the cyclotron resonance appear naturally as a consequence of the nonlocal response of the 2DES. Combining Eqs. (2) and (3), and the conductivities with Ohm's law, we obtain the self-consistent current-density equations

$$j_x(q_x, \omega) = -\frac{i\kappa}{2\varepsilon\omega}\sigma_{xx}(q_x, \omega)j_x(q_x, \omega) + \frac{i\mu\omega}{2\kappa}\sigma_{xy}(q_x, \omega)j_y(q_x, \omega), \tag{13}$$

and

$$j_y(q_x, \omega) = -\frac{i\kappa}{2\varepsilon\omega}\sigma_{yx}(q_x, \omega)j_x(q_x, \omega) + \frac{i\mu\omega}{2\kappa}\sigma_{yy}(q_x, \omega)j_y(q_x, \omega). \tag{14}$$

Requiring a nontrivial solution for $(j_x, j_y)$ yields the dispersion equation

$$\Lambda_{\text{full}}(q_x, \omega) = 0, \tag{15}$$

where

$$\Lambda_{full}(q_x,\omega) = \left[1 + \frac{i\sqrt{q_x^2 - \varepsilon\mu\omega^2}}{2\varepsilon\omega}\sigma_{xx}(q_x,\omega)\right]\left[1 - \frac{i\mu\omega}{2\sqrt{q_x^2 - \varepsilon\mu\omega^2}}\sigma_{yy}(q_x,\omega)\right] - \frac{\mu}{4\varepsilon}\sigma_{xy}(q_x,\omega)\sigma_{yx}(q_x,\omega). \quad (16)$$

The solutions of Eq. (15) describe the collective eigenmodes of the 2DES, including the magnetoplasmon mode and its hybridization with the higher harmonics of the cyclotron resonance. A similar expression has been reported in Ref. [4].

# III. Finite-wave-vector activation of higher cyclotron harmonics

We first analyze the eigenmodes by calculating $|\Lambda_{\mathrm{full}}(q_x,\omega)|$ as a function of magnetic field $B$ and frequency. Assuming that the 2DES is embedded in a homogeneous GaAs medium, we used the magnetic permeability $\mu = \mu_0$ and the permittivity $\varepsilon_{\mathrm{GaAs}} = 12.9\varepsilon_0$. The electron effective mass $m^*$ and the 2DES temperature are set to $m^* = 0.067m_0$ and 0 K. It should be noted that the scattering time of electrons is phenomenologically considered by replacing $\omega + (n-m)\omega_c$ by $\omega + (n-m)\omega_c + i/\tau$ with the scattering time of $\tau = 40$ ps.

Figures 2(a)–2(d) show $|\Lambda_{\mathrm{full}}(q_x,\omega)|$ calculated for different in-plane wavelength $2\pi/q_x =$ 1 μm, 0.5 μm, 0.1 μm, and 0.05 μm, respectively. The dark regions in the color maps correspond to minima in $|\Lambda_{\mathrm{full}}(q_x,\omega)|$, identifying the collective-mode dispersions. In Fig. 2(a), the mode labeled MP is assigned to the magnetoplasmon mode. The mode labeled 2CR follows approximately $2\omega_c$ and is therefore attributed to the second harmonic of the cyclotron resonance. An avoided crossing opens near the intersection of the MP and 2CR modes, indicating hybridization between the collective magnetoplasmon excitation and the finite-$q_x$-induced second harmonic of the cyclotron resonance. As the wave vector $q_x$ increases, the magnetoplasmon frequency shifts upward. At the same time, the avoided-crossing gap between the MP and 2CR modes becomes larger, indicating that the coupling strength is enhanced at larger wave vector. For $2\pi/q_x = 0.5$ μm and 0.1 μm in Figs. 2(b) and 2(c), the third harmonic of the cyclotron resonance is also visible, and an additional avoided crossing appears near the intersection between the MP and 3CR modes. For $2\pi/q_x = 0.05$ μm in Fig. 2(d), avoided crossings involving higher-order CR modes become substantially more pronounced. These results show that finite-wave-vector near fields provide a systematic way to activate and hybridize higher cyclotron harmonics with magnetoplasmon modes.

We next examine the electron-density dependence of the MP–nCR coupled system. Figures 3(a)–3(d) show $|\Lambda_{\mathrm{full}}(q_x,\omega)|$ calculated for $n_s = 1.0\times10^{14}$, $1.0\times10^{15}$, $1.0\times10^{16}$, and $1.0\times10^{17}$ $\mathrm{m}^{-2}$, respectively, with the wavelength fixed at $2\pi/q_x = 0.1$ μm. For Figs. 3(b)-3(d), the plasmon frequency increases with increasing electron density.

# IV. Analytical expression for the effective coupling strength

To clarify the dependence of the coupling strength between the MP and the nCR on the wavelength $2\pi/q_x$ and electron density $n_s$, we derive an approximate analytical expression for the coupling strength. In the nonretarded limit, $q_x^2 \gg \varepsilon\mu\omega^2$, and neglecting dissipation, the dispersion function reduces to

$$\Lambda_{full}(q_x,\omega) \simeq 1 + i\frac{q_x}{2\varepsilon\omega}\sigma_{xx} \equiv \Lambda_{approx}(q_x,\omega). \tag{17}$$

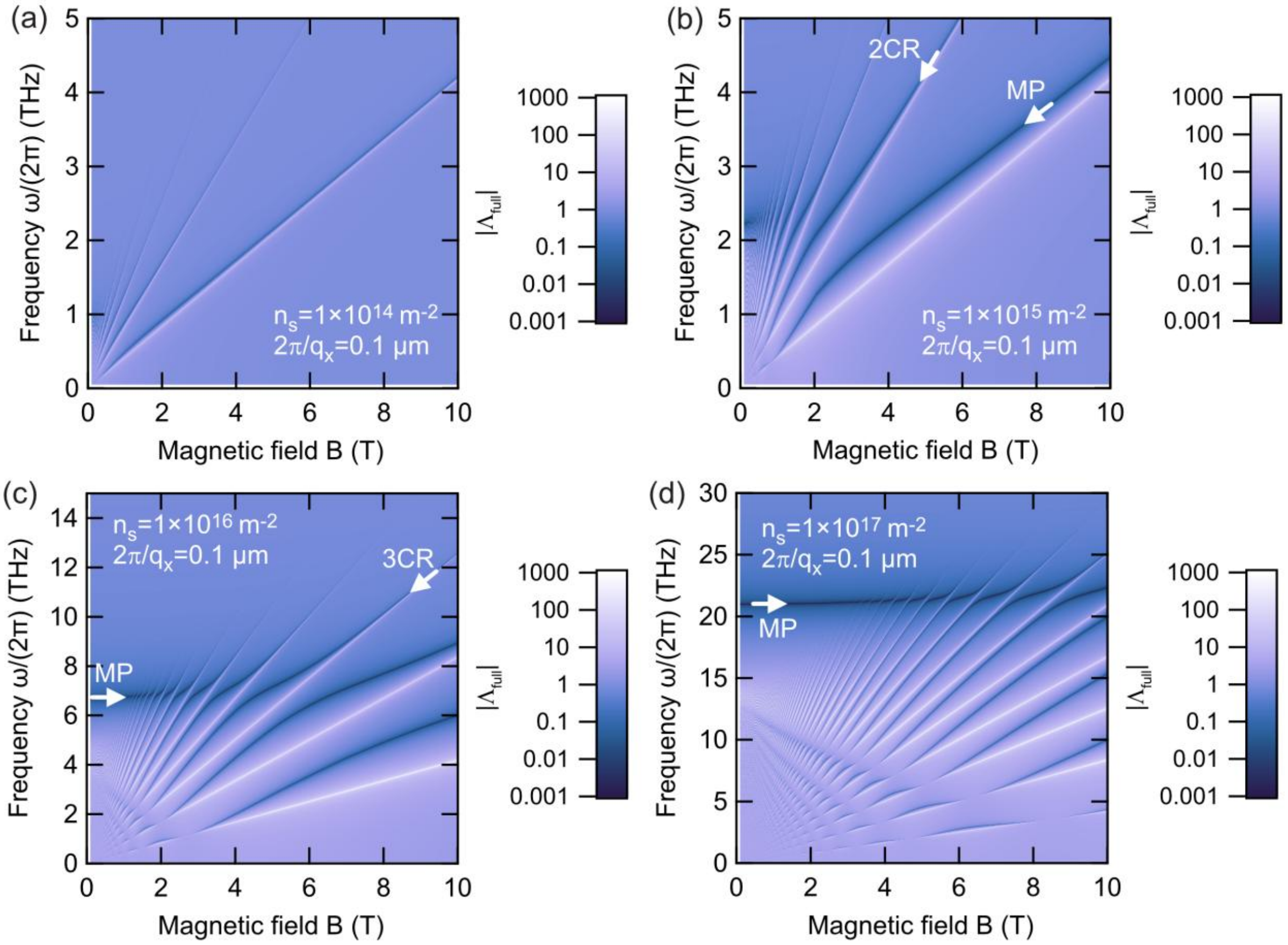

Fig. 3. Color-coded maps of $|\Lambda_{\mathrm{full}}(q_x,\omega)|$ for the different values of $n_s$ (the values are written in the plots) at the fixed wavelength $2\pi/q_x = 0.1$ μm and at the 2DES temperature of 0 K. The dark regions in the color maps correspond to the solutions of the dispersion equation in Eq. (16).

The $xx$ component of the conductivity in Eq. (10) is rewritten as

$$\sigma_{xx}(q_x,\omega) = i\frac{e^2 n_s}{m^*}\sum_{n=1}^{\infty}\mathcal{F}(q_x,n)\left[\frac{\omega}{\omega^2-(n\omega_c)^2}\right], \tag{18}$$

where

$$\mathcal{F}(q_x,n) = \frac{2m^*\omega_c}{\pi\hbar l_B^2 n_s}\left(\frac{n}{q_x^2}\right)\sum_{m=0}^{\infty}(f_m - f_{m+n})F_{m,m+n}(q_x)F_{m+n,m}(-q_x). \tag{19}$$

The dispersion relation becomes

$$1 - \frac{e^2 n_s q_x}{2\varepsilon m^*}\sum_{n=1}^{\infty}\frac{\mathcal{F}(q_x,n)}{\omega^2-(n\omega_c)^2} = 0. \tag{20}$$

This expression can be further modified as

$$\frac{\omega_p^2 + \omega_c^2}{\omega^2} = 1 - \frac{1}{\omega^2} \sum_{n=2}^{\infty} \frac{(n^2 - 1)\omega_c^2 \omega_p^2}{\omega^2 - (n\omega_c)^2} \mathcal{F}(q_x, n), \tag{21}$$

where

$$\begin{aligned} \omega_p^2 &= \frac{e^2 n_s q_x}{2\varepsilon m^*} \sum_{n=1}^{\infty} \mathcal{F}(q_x, n) \\ &= \frac{e^2 n_s q_x}{2\varepsilon m^*}. \end{aligned} \tag{22}$$

is a well-known expression of the bare plasmon frequency within the local approximation. The second equation is given by the oscillator-strength sum rule (see Appendix B). Equation (21) shows that higher cyclotron harmonics appear as oscillator-like resonances coupled to the magnetoplasmon mode in the collective-mode dispersion relation. This approximate dispersion relation accurately reproduces the eigenfrequencies obtained from the full dispersion equation in Eq. (16) (see Appendix D).

Next, we map the Maxwell–Kubo dispersion relation onto a Hopfield-type Hamiltonian [27]. The Hopfield Hamiltonian with a diamagnetic term is

$$\hat{\mathcal{H}}/\hbar = \omega_{\mathrm{MP}} \hat{a}^\dagger \hat{a} + \sum_{n=2} \left[ n\omega_c \hat{b}_n{}^\dagger \hat{b}_n + i g_n (\hat{a} + \hat{a}^\dagger) \left( \hat{b}_n - \hat{b}_n{}^\dagger \right) \right] + D(\hat{a} + \hat{a}^\dagger)^2. \tag{23}$$

Here, $\hat{a}$ is the annihilation operator of the magnetoplasmon mode, and $\hat{b}_n$ is the collective annihilation operator of the $n$-th cyclotron-resonance harmonic. The parameter $g_n$ denotes the effective coupling strength between the magnetoplasmon mode and the $n$-th harmonic of the cyclotron resonance, while $D$ represents the diamagnetic term. The dispersion relation derived from the Hopfield Hamiltonian is given by

$$\frac{\omega_{\mathrm{MP}}^2}{\omega^2} = 1 - \frac{1}{\omega^2} \left[ 4\omega_{\mathrm{MP}} D + \sum_{n=2} \frac{4 g_n^2 n\omega_c \omega_{\mathrm{MP}}}{\omega^2 - (n\omega_c)^2} \right]. \tag{24}$$

The detailed derivation of Eq. (24) for the Hopfield Hamiltonian is given in Appendix C. By comparing Eqs. (21) and (24), we find the expressions for the coupling strength as

$$g_n^2 = \frac{n^2 - 1}{4n^2} \frac{n\omega_c \omega_p^2}{\omega_{\mathrm{MP}}} \mathcal{F}(q_x, n). \tag{25}$$

Within the present effective parametrization, comparison of Eqs. (21) and (24) gives $D = 0$. At the resonance condition $\omega_{\mathrm{MP}} = n\omega_c$, we have $\omega_p^2 = (n^2 - 1)\omega_c^2$. The normalized coupling strength at the resonance is then

$$\frac{g_n^{\mathrm{res}}}{\omega_p} = \frac{\sqrt{n^2 - 1}}{2n} \sqrt{\mathcal{F}(q_x, n)}. \tag{26}$$

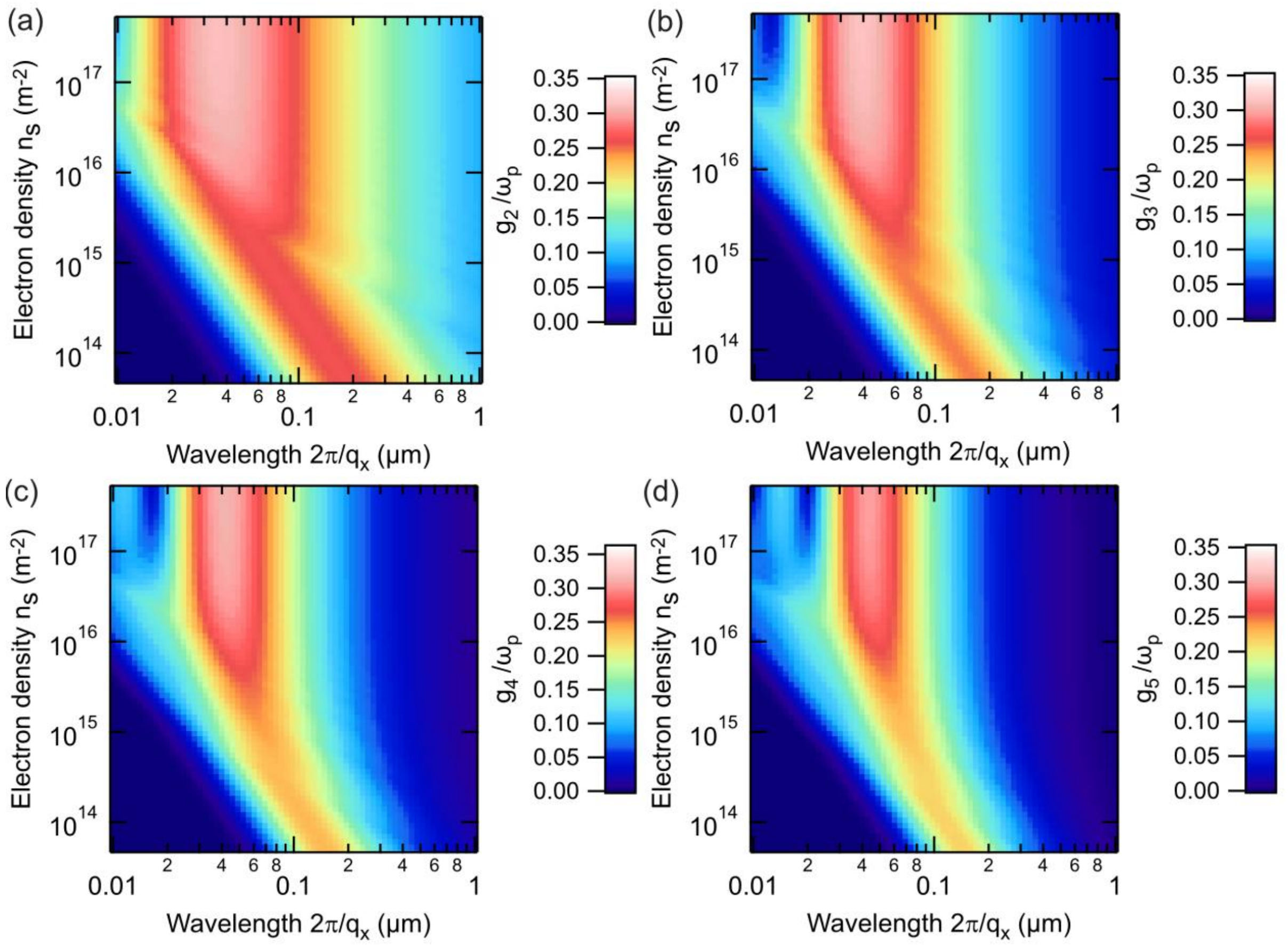

Fig. 4. The normalized coupling strength $g_n^{\mathrm{res}}/\omega_p$ as a function of the wavelength $2\pi/q_x$ and the electron density $n_s$ for the different harmonics $n$ of the CR.

This expression is one of the central results of this work. Figures 4(a)–4(d) show the normalized coupling strength $g_n^{\mathrm{res}}/\omega_p$ as a function of the wavelength $2\pi/q_x$ and the electron density $n_s$ for the different harmonics $n$ of the CR. The ultrastrong-coupling regime is commonly characterized by $g_n^{\mathrm{res}}/\omega_p > 0.1$. In Fig. 4(a), for $n_s \lesssim 10^{16}\ \mathrm{m}^{-2}$, the wavelength at which the normalized coupling strength reaches its maximum shifts toward shorter wavelengths with increasing $n_s$. In contrast, for $n_s \gtrsim 10^{16}\ \mathrm{m}^{-2}$, the peak wavelength becomes nearly independent of $n_s$, remaining at approximately 40 nm. In this high-density regime, the maximum normalized coupling strength approaches approximately 0.3 and becomes nearly independent of $n_s$. Furthermore, the maximum value is nearly independent of material parameters, such as the effective mass $m^*$ and the effective permittivity $\varepsilon$, although the optimal wavelength depends on these parameters. Similar trends are observed for higher-order harmonics of the CR, suggesting a universal upper limit on the normalized coupling strength for each cyclotron harmonic.

# V. Conclusion

We have theoretically investigated the finite-wave-vector magnetoplasmon response of a two-dimensional electron system in a perpendicular magnetic field. By combining Maxwell's equations with the nonlocal conductivity tensor obtained from the Kubo formula, we derived a self-consistent dispersion relation for the collective modes. The finite in-plane wave vector allows higher-order cyclotron transitions, which are forbidden in the long-wavelength limit, to contribute to the nonlocal magnetoconductivity. As a result, the magnetoplasmon mode hybridizes with the higher cyclotron harmonics, giving rise to avoided crossings in the collective-mode spectrum.

We established a correspondence between the Maxwell–Kubo dispersion relation and a Hopfield-type coupled-oscillator Hamiltonian describing the interaction between the magnetoplasmon and higher-order cyclotron excitations. This mapping provides a quantitative definition of the effective coupling strength between the magnetoplasmon and the $n$-th cyclotron-resonance harmonic.

Our analysis reveals that the normalized coupling strength can be enhanced by increasing the in-plane wave vector, but reaches a maximum at a characteristic wavelength. For the second harmonic, the peak wavelength approaches approximately 40 nm in the high-electron-density regime, where the maximum normalized coupling strength reaches approximately 0.3. Similar behavior is found for higher-order cyclotron harmonics. These findings provide a quantitative guideline for optimizing the coupling between magnetoplasmons and higher cyclotron harmonics through the spatial confinement of electromagnetic fields.

Our theoretical framework provides a basis for designing THz magnetoplasmonic systems with enhanced coupling strengths, potentially reaching the ultrastrong-coupling regime. Furthermore, because finite-wave-vector electromagnetic fields can probe internal electronic dynamics beyond the uniform center-of-mass response constrained by Kohn's theorem, this approach may open opportunities for investigating the influence of electron–electron interactions on strongly coupled collective excitations.

# Acknowledgement

This work was supported by JSPS KAKENHI Grant Numbers JP26K01332 (K.K.), JP25K01694 (M.B.), JP25K01691 (M.B.), JP25K00012 (M.B.) JP24K21526 (M.B.), JP22K03481 (K.K.), JST PRESTO Grant Number JPMJPR2255 (K.K.), JST FOREST Grant Number JPMJFR246Q (K.K.), JST CREST Grant Number JPMJCR25A3 (K.K.), General Incorporated Association Kyoso (K.K.), the Sumitomo Foundation (K.K.), and Murata Science and Education Foundation (K.K.).

# Appendix

## Appendix A: Electromagnetic boundary condition for a two-dimensional sheet current

In this Appendix, we derive Eqs. (2) and (3) in the main text. We define the Fourier transform of the electric field as

$$\mathbf{E}(x,z,t)=\left(\frac{1}{2\pi}\right)^2\int dq_x d\omega\,\mathbf{E}(q_x,z,\omega)e^{i(q_x x-\omega t)}. \qquad (A1)$$

We substitute the Fourier transform in Eq. (A1) into the wave equation

$$\nabla \times [\nabla \times \mathbf{E}(\mathbf{r},t)] + \varepsilon\mu \frac{\partial^2}{\partial t^2}\mathbf{E}(\mathbf{r},t) = -\mu \frac{\partial}{\partial t}[\mathbf{j}(x,t)\delta(z)]. \qquad (A2)$$

Using the vector identity $\nabla \times (\nabla \times \boldsymbol{A}) = \nabla(\nabla \cdot \boldsymbol{A}) - \nabla^2 \boldsymbol{A}$, we obtain

$$-\frac{\partial^2 E_x}{\partial z^2} + iq_x \frac{\partial E_z}{\partial z} - \varepsilon\mu\omega^2 E_x = i\omega\mu j_x \delta(z), \qquad (A3)$$

$$-\frac{\partial^2 E_y}{\partial z^2} + (q_x^2 - \varepsilon\mu\omega^2)E_y = i\omega\mu j_y \delta(z), \qquad (A4)$$

$$iq_x \frac{\partial E_x}{\partial z} + (q_x^2 - \varepsilon\mu\omega^2)E_z = 0. \qquad (A5)$$

Eliminating $E_z$ from Eqs. (A3) and (A5), we obtain

$$-\frac{\partial^2 E_x}{\partial z^2} + (q_x^2 - \varepsilon\mu\omega^2)E_x = -i\frac{q_x^2 - \varepsilon\mu\omega^2}{\varepsilon\omega} j_x \delta(z). \qquad (A6)$$

For $z \neq 0$, the solution is

$$E_x(q_x, z, \omega) = E_x(q_x, 0, \omega)e^{-\kappa|z|}, \qquad (A7)$$

where

$$\kappa(q_x, \omega) \equiv \sqrt{q_x^2 - \varepsilon\mu\omega^2}, \qquad (A8)$$

with Re $\kappa > 0$. Integrating Eq. (A6) across z = 0 gives

$$E_x(q_x, 0, \omega) = -\frac{i\kappa}{2\varepsilon\omega} j_x(q_x, \omega), \qquad (A9)$$

where the Fourier component of the $x$ component of the current density is defined as

$$j_x(x,t) = \left(\frac{1}{2\pi}\right)^2 \int dq_x d\omega \; j_x(q_x, \omega)e^{i(q_x x - \omega t)}. \qquad (A10)$$

A similar calculation for the $y$ component gives

$$E_y(q_x, 0, \omega) = \frac{i\mu\omega}{2\kappa} j_y(q_x, \omega). \qquad (A11)$$

## Appendix B: Landau-level Kubo formulation of the nonlocal magnetoconductivity tensor

### Hamiltonian and Landau levels

We consider a two-dimensional electron system (2DES) in a perpendicular magnetic field $B$. We adopt the Landau gauge, in which the vector potential for the static magnetic field is given by

$$\mathbf{A}_0(\mathbf{r}) = (0, Bx, 0). \qquad (B1)$$

The single-particle Hamiltonian of an electron in the 2DES is

$$h_0 = \frac{(\mathbf{p} + e\mathbf{A}_0)^2}{2m^*}$$
$$= \frac{p_x^2 + \left(p_y + eBx\right)^2}{2m^*}$$
$$= \frac{\mathbf{\Pi}^2}{2m^*}, \tag{B2}$$

where the kinetic momentum $\mathbf{\Pi} = \left(\Pi_x, \Pi_y\right)$ is defined as

$$\Pi_x = p_x, \tag{B3}$$
$$\Pi_y = p_y + eBx, \tag{B4}$$

and $m^*$is the electron effective mass. Here, $e$ denotes the elementary charge, and the electron charge is $-e$. The eigenstate of this Hamiltonian is written as

$$\psi_{nk}(x, y) = \frac{1}{\sqrt{L_y}} e^{iky} \phi_n(x - X_k), \tag{B5}$$

where the guiding-center coordinate is

$$X_k = -kl_B^2, \tag{B6}$$

and the magnetic length is

$$l_B = \sqrt{\frac{\hbar}{eB}}. \tag{B7}$$

The harmonic-oscillator wavefunction $\phi_n(x)$ is given by

$$\phi_n(x) = \frac{1}{\sqrt{2^n n!}} \frac{1}{\pi^{1/4}\sqrt{l_B}} H_n\left(\frac{x}{l_B}\right) \exp\left(-\frac{x^2}{2l_B^2}\right), \tag{B8}$$

where $H_n(x)$ is the Hermite polynomial. The corresponding eigenenergy is

$$\varepsilon_n = \hbar\omega_c\left(n + \frac{1}{2}\right). \tag{B9}$$

Finally, we define the ladder operator for the landau levels as

$$c = \frac{l_B}{\sqrt{2}\hbar}\left(\Pi_y + i\Pi_x\right), \tag{B10}$$
$$c^\dagger = \frac{l_B}{\sqrt{2}\hbar}\left(\Pi_y - i\Pi_x\right), \tag{B11}$$

where $[c, c^\dagger] = 1$ is satisfied. Using the ladder operator, the kinetic momentum can be expressed as

$$\Pi_x = -i\frac{\hbar}{\sqrt{2}l_B}(c - c^\dagger), \tag{B12}$$
$$\Pi_y = \frac{\hbar}{\sqrt{2}l_B}(c + c^\dagger). \tag{B13}$$

**Perturbative vector potential with wave vector and frequency**

We next introduce an external electromagnetic field with a finite wave vector and frequency as a small perturbation. The total vector potential is written as

$$\mathbf{A}_{\text{tot}}(\mathbf{r},t)=\mathbf{A}_0(\mathbf{r})+\mathbf{A}(\mathbf{r},t). \quad (B14)$$

The single-particle Hamiltonian under the total vector potential is

$$\begin{aligned} h &= \frac{(\mathbf{p}+e\mathbf{A}_{\text{tot}})^2}{2m^*} \\ &= \frac{\mathbf{\Pi}^2}{2m^*}+\frac{\mathrm{e}[\mathbf{\Pi}\mathbf{A}(\mathbf{r},t)+\mathbf{A}(\mathbf{r},t)\mathbf{\Pi}]}{2m^*}+\frac{e^2\mathbf{A}^2}{2m^*}, \end{aligned} \quad (B15)$$

where

$$\mathbf{\Pi}=\mathbf{p}+\mathrm{e}\mathbf{A}_0(\mathbf{r}), \quad (B16)$$

is the kinetic momentum in the static magnetic field. The Hamiltonian can be decomposed as

$$h=h_0+h_p+h_d, \quad (B17)$$

where

$$h_p=\frac{e[\mathbf{\Pi}\cdot\mathbf{A}(\mathbf{r},t)+\mathbf{A}(\mathbf{r},t)\cdot\mathbf{\Pi}]}{2m^*}, \quad (B18)$$

$$h_d=\frac{e^2}{2m^*}\mathbf{A}^2(\mathbf{r},t). \quad (B19)$$

Here, $h_p$ is the term linear in the perturbative vector potential, and $h_d$ is the diamagnetic term quadratic in $\mathbf{A}$. For the many-body electron system, the corresponding perturbation Hamiltonians are

$$\mathcal{H}_p=\sum_i\frac{e[\mathbf{\Pi}_\mathrm{i}\cdot\mathbf{A}(\mathbf{r}_i,t)+\mathbf{A}(\mathbf{r}_i,t)\cdot\mathbf{\Pi}_\mathrm{i}]}{2m^*}, \quad (B20)$$

$$\mathcal{H}_d=\sum_i\frac{e^2}{2m^*}\mathbf{A}^2(\mathbf{r}_i,t). \quad (B21)$$

Here, $\hat{\mathbf{r}}_i$ and $\mathbf{\Pi}_\mathrm{i}$ denote the position and kinetic momentum operators of the $i$-th electron.

**Paramagnetic current response**

First, we consider the current response induced by the term linear in the perturbative vector potential. We define the paramagnetic current-density operator with respect to the perturbative vector potential as

$$\hat{\mathbf{j}}^p(\mathbf{r})=-\frac{e}{2m^*}\sum_i\left[\widehat{\mathbf{\Pi}}_i\delta(\mathbf{r}-\hat{\mathbf{r}}_i)+\delta(\mathbf{r}-\hat{\mathbf{r}}_i)\widehat{\mathbf{\Pi}}_i\right]. \quad (B22)$$

With this definition, the perturbation Hamiltonian linear in **A** is written as

$$\mathcal{H}_p = -\int d^2r\, \hat{\mathbf{j}}^p(\mathbf{r}) \cdot \mathbf{A}(\mathbf{r}, t). \tag{B23}$$

The retarded paramagnetic current–current response function is defined by

$$\chi_{j_\alpha^p j_\beta^p}(\mathbf{r}, \mathbf{r}', t) = -\frac{i}{\hbar}\theta(t)\langle\left[j_\alpha^p(\mathbf{r}, t), j_\beta^p(\mathbf{r}')\right]\rangle_0, \tag{B24}$$

where $\alpha, \beta = x, y$ denote Cartesian components. The Heisenberg representation of the current-density operator is

$$j_\alpha^p(\mathbf{r}, t) = e^{\frac{i}{\hbar}\mathcal{H}_0 t} j_\alpha^p(\mathbf{r}) e^{-\frac{i}{\hbar}\mathcal{H}_0 t}. \tag{B25}$$

Here, $\langle\cdots\rangle_0$ denotes the thermal-equilibrium average with respect to the unperturbed Hamiltonian $H_0$. The Fourier transform of the paramagnetic current-density operator is defined as

$$\begin{aligned}\hat{\mathbf{j}}^p(\mathbf{q}) &= \int d^2r\, e^{-i\mathbf{q}\cdot\mathbf{r}}\, \hat{\mathbf{j}}^p(\mathbf{r}) \\ &= -\frac{e}{2m^*}\sum_i \left[\widehat{\mathbf{\Pi}}_i e^{-i\mathbf{q}\cdot\hat{\mathbf{r}}_i} + e^{-i\mathbf{q}\cdot\hat{\mathbf{r}}_i}\widehat{\mathbf{\Pi}}_i\right].\end{aligned} \tag{B26}$$

The Fourier transform of the retarded current–current response function is defined by

$$\chi_{j_\alpha^p j_\beta^p}(\mathbf{q}, \mathbf{q}', \omega) = \frac{1}{S}\int d^2r\, e^{-i\mathbf{q}\cdot\mathbf{r}} \int d^2r'\, e^{i\mathbf{q}'\cdot\mathbf{r}'} \int_0^\infty dt\, e^{i\omega t}\chi_{j_\alpha^p j_\beta^p}(\mathbf{r}, \mathbf{r}', t), \tag{B27}$$

where $S = L_x L_y$ is the area of the 2DES. The paramagnetic current density can be written as

$$\langle \hat{j}_\alpha^p(\mathbf{q}, \omega)\rangle = -\sum_\beta \sum_{\mathbf{q}'} \chi_{j_\alpha^p j_\beta^p}(\mathbf{q}, \mathbf{q}', \omega) A_\beta(\mathbf{q}', \omega). \tag{B28}$$

For a homogeneous 2DES in a uniform magnetic field, the physical electromagnetic response is translationally invariant after the Landau-level degeneracy is properly summed over. After summing over the guiding-center degeneracy, the response depends only on $(\mathbf{r} - \mathbf{r}')$. As a result, the Fourier-transformed response function becomes diagonal with respect to wave vector:

$$\chi_{j_\alpha^p j_\beta^p}(\mathbf{q}, \mathbf{q}', \omega) = \chi_{j_\alpha^p j_\beta^p}(\mathbf{q}, \omega)\delta_{\mathbf{q},\mathbf{q}'}. \tag{B29}$$

Therefore, the expression of the paramagnetic current density reduces to

$$\langle \hat{j}_\alpha^p(\mathbf{q}, \omega)\rangle = -\sum_\beta \chi_{j_\alpha^p j_\beta^p}(\mathbf{q}, \omega) A_\beta(\mathbf{q}, \omega) \tag{B30}$$

**Diamagnetic current response**

The physical current consists of the diamagnetic contribution as well as the paramagnetic contribution. The number-density operator is defined as

$$\hat{n}(\mathbf{r}) = \sum_i \delta(\mathbf{r} - \hat{\mathbf{r}}_i)\,. \tag{B31}$$

The diamagnetic current-density operator is obtained from the diamagnetic term $H_d$. It is given by

$$\hat{\mathbf{j}}^d(\mathbf{r}, t) = -\frac{e^2}{m^*}\hat{n}(\mathbf{r})\mathbf{A}(\mathbf{r}, t). \tag{B32}$$

Indeed, the diamagnetic Hamiltonian can be written as

$$\begin{aligned}\mathcal{H}_d &= \int d^2r \sum_i \frac{e^2}{2m^*}\delta(\mathbf{r} - \hat{\mathbf{r}}_i)\,\mathbf{A}^2(\mathbf{r}, t) \\ &= -\frac{1}{2}\int d^2r\,\hat{\mathbf{j}}^d(\mathbf{r}, t)\cdot\mathbf{A}(\mathbf{r}, t).\end{aligned} \tag{B33}$$

In the linear-response regime, the density operator in the diamagnetic current can be replaced by the equilibrium electron density $n_s(\mathbf{r})$. Thus,

$$\langle\hat{\mathbf{j}}^d(\mathbf{r}, t)\rangle \simeq -\frac{e^2}{m^*}n_s(\mathbf{r})\mathbf{A}(\mathbf{r}, t). \tag{B34}$$

Defining the diamagnetic response kernel through

$$\langle\hat{j}^d_\alpha(\mathbf{r}, t)\rangle = \sum_\beta \int d^2r' \int dt'\,K^d_{\alpha\beta}(\mathbf{r}, \mathbf{r}', t, t')A_\beta(\mathbf{r}', t')\,, \tag{B35}$$

the diamagnetic contribution to the current response kernel is an instantaneous contact term:

$$K^d_{\alpha\beta}(\mathbf{r}, \mathbf{r}', t, t') = -\frac{e^2}{m^*}n_s(\mathbf{r})\delta_{\alpha\beta}\delta(\mathbf{r} - \mathbf{r}')\delta(t - t'). \tag{B36}$$

For a uniform electron density $n_s$, its Fourier transform becomes

$$K^d_{\alpha\beta}(\mathbf{q}, \omega) = -\frac{e^2 n_s}{m^*}\delta_{\alpha\beta}, \tag{B37}$$

and that of the current density can be written as

$$\langle\hat{j}^d_\alpha(\mathbf{q}, \omega)\rangle = \sum_\beta K^d_{\alpha\beta}(\mathbf{q}, \omega)A_\beta(\mathbf{q}, \omega)\,. \tag{B38}$$

**Kubo electrical conductivity in the Landau-level basis**

The total current density is the sum of the paramagnetic and diamagnetic contributions:

$$\langle\hat{j}_\alpha(\mathbf{q}, \omega)\rangle = \sum_\beta \left[-\chi_{j^p_\alpha j^p_\beta}(\mathbf{q}, \omega) + K^d_{\alpha\beta}(\mathbf{q}, \omega)\right]A_\beta(\mathbf{q}, \omega)\,. \tag{B39}$$

The total current-response kernel is given by

$$K_{\alpha\beta}(\mathbf{q},\omega) = -\chi_{j_\alpha^p j_\beta^p}(\mathbf{q},\omega) - \frac{e^2 n_0}{m^*}\delta_{\alpha\beta}. \tag{B40}$$

Here, the minus sign in front of the paramagnetic response function follows from the sign convention used in Eq. (B24), together with the perturbation Hamiltonian $\mathcal{H}_p$. Using the time dependence $e^{-i\omega t}$, the electric field and vector potential are related by

$$\mathbf{E}(\mathbf{q},\omega) = i\omega\mathbf{A}(\mathbf{q},\omega). \tag{B41}$$

Therefore, the finite-wave-vector conductivity tensor is obtained as

$$\begin{aligned}\sigma_{\alpha\beta}(q,\omega) &= \frac{K_{\alpha\beta}(\mathbf{q},\omega)}{i\omega} \\ &= \frac{i}{\omega}\left[\frac{e^2 n_0}{m^*}\delta_{\alpha\beta} + \chi_{j_\alpha^p j_\beta^p}(\mathbf{q},\omega)\right].\end{aligned} \tag{B42}$$

We now evaluate the paramagnetic current–current response function using the Landau-level basis. For noninteracting electrons, the Lehmann representation of the response function for Landau levels is

$$\begin{aligned}\chi_{j_\alpha^p j_\beta^p}(\mathbf{q},\omega) &= \frac{1}{S}\sum_{n,m}\sum_{k,k'}\frac{f_n - f_m}{\hbar\omega + \varepsilon_n - \varepsilon_m}\langle nk|j_\alpha^p(\mathbf{q})|mk'\rangle\left\langle mk'\middle|j_\beta^p(-\mathbf{q})\middle|nk\right\rangle \\ &= \frac{1}{S}\sum_{n,m}\sum_{k,k'}\frac{f_n - f_m}{\hbar\omega + (n-m)\hbar\omega_c}\langle nk|j_\alpha^p(\mathbf{q})|mk'\rangle\left\langle mk'\middle|j_\beta^p(-\mathbf{q})\middle|nk\right\rangle,\end{aligned} \tag{B43}$$

where the paramagnetic current-density operator in momentum space is

$$\mathbf{j}^p(\mathbf{q}) = -\frac{e}{2m^*}\left(\mathbf{\Pi}e^{-i\mathbf{q}\cdot\mathbf{r}} + e^{-i\mathbf{q}\cdot\mathbf{r}}\mathbf{\Pi}\right). \tag{B44}$$

Here, $f_n$ is the occupation factor of the $n$-th Landau level. For a finite wave vector along the $x$ direction, $\mathbf{q} = q_x\mathbf{e}_x$, the matrix element of the $x$ component of the paramagnetic current density can be calculated as

$$\langle nk|j_x^p(q_x)|mk'\rangle = e\omega_c\left(\frac{n-m}{q_x}\right)F_{nm}(q_x)e^{-iq_x X_k}\delta_{k,k'}, \tag{B45}$$

where we introduced the Landau-level form factor defined as

$$F_{nm}(q_x) = \int du\, e^{-iq_x l_B u}\varphi_n(u)\varphi_m(u). \tag{B46}$$

Here, we introduce the dimensionless coordinate

$$u = \frac{x - X_k}{l_B}, \tag{B47}$$

and the harmonic-oscillator wavefunction $\phi_n(x - X_k)$ is written as

$$\phi_n(x - X_k) = \frac{1}{\sqrt{2^n n!}} \frac{1}{\pi^{1/4}\sqrt{l_B}} H_n(u) \exp\left(-\frac{u^2}{2}\right)$$
$$\equiv \frac{1}{\sqrt{l_B}} \varphi_n(u). \quad (B48)$$

Similarly, the $y$ component of the current matrix element becomes

$$\langle nk | j_y^p(q_x) | mk' \rangle = -ie\omega_c \left[\frac{\partial}{\partial q_x} F_{nm}(q_x)\right] e^{-iq_x X_k} \delta_{k,k'}. \quad (B49)$$

The retarded paramagnetic current–current response function for Landau levels can be rewritten in terms of the current matrix elements derived above. For the $xx$ component, we obtain

$$\chi_{j_x^p j_x^p}(q_x, \omega) = \frac{e^2\omega_c^2}{\pi\hbar l_B^2} \sum_{n,m} \frac{f_n - f_m}{\omega + (n-m)\omega_c} \left(\frac{n-m}{q_x}\right)^2 F_{nm}(q_x) F_{mn}(-q_x), \quad (B50)$$

where we take into account the degeneracy of the Landau level with respect to the wave vector $k$ as

$$\frac{1}{S}\sum_{k,k'} \delta_{k,k'} = \frac{1}{S}\sum_k 1 = 2 \times \frac{1}{2\pi l_B^2}. \quad (B51)$$

Here, the prefactor of 2 acconuts for the spin degeneracy. Similarly, the $xy$ component is given by

$$\chi_{j_x^p j_y^p}(q_x, \omega) = i \frac{e^2\omega_c^2}{\pi\hbar l_B^2} \sum_{n,m} \frac{f_n - f_m}{\omega + (n-m)\omega_c} \left(\frac{n-m}{q_x}\right) F_{nm}(q_x) \frac{\partial F_{mn}(-q_x)}{\partial q_x}. \quad (B52)$$

Here, the derivative $\partial F_{mn}(-q_x)/\partial q_x$ means the derivative of the entire function $F_{mn}(-q_x)$ with respect to $q_x$. The conductivity tensor is obtained by adding the diamagnetic contribution. For an electron sheet density $n_s$ and with the convention $\mathbf{E}(q_x, \omega) = i\omega \mathbf{A}(q_x, \omega)$, we have

$$\sigma_{xx}(q_x, \omega) = \frac{i}{\omega}\left[\frac{e^2 n_s}{m^*} + \chi_{j_x^p j_x^p}(q_x, \omega)\right]$$
$$= \frac{i}{\omega}\left[\frac{e^2 n_s}{m^*} + \frac{e^2\omega_c^2}{\pi\hbar l_B^2} \sum_{n,m} \frac{f_n - f_m}{\omega + (n-m)\omega_c} \left(\frac{n-m}{q_x}\right)^2 F_{nm}(q_x) F_{mn}(-q_x)\right]. \quad (B53)$$

For the Hall component, there is no diamagnetic contribution. Thus,

$$\sigma_{xy}(q_x, \omega) = \frac{i}{\omega} \chi_{j_x^p j_y^p}(q_x, \omega)$$
$$= -\frac{e^2\omega_c^2}{\pi\hbar\omega l_B^2} \sum_{n,m} \frac{f_n - f_m}{\omega + (n-m)\omega_c} \left(\frac{n-m}{q_x}\right) F_{nm}(q_x) \frac{\partial F_{mn}(-q_x)}{\partial q_x}. \quad (B54)$$

Similarly, we obtain the $yx$ component

$$\sigma_{yx}(q_x,\omega)=\frac{i}{\omega}\chi_{j_y^p j_x^p}(q_x,\omega)$$
$$=\frac{e^2\omega_c^2}{\pi\hbar\omega l_B^2}\sum_{n,m}\frac{f_n-f_m}{\omega+(n-m)\omega_c}\left(\frac{n-m}{q_x}\right)\frac{\partial F_{nm}(q_x)}{\partial q_x}F_{mn}(-q_x)\,, \qquad (B55)$$

and the $yy$ component

$$\sigma_{yy}(q_x,\omega)=\frac{i}{\omega}\left[\frac{e^2n_s}{m^*}+\chi_{j_y^p j_y^p}(q_x,\omega)\right]$$
$$=\frac{i}{\omega}\left\{\frac{e^2n_s}{m^*}+\frac{e^2\omega_c^2}{\pi\hbar l_B^2}\sum_{n,m}\frac{f_n-f_m}{\omega+(n-m)\omega_c}\frac{\partial F_{nm}(q_x)}{\partial q_x}\frac{\partial F_{mn}(-q_x)}{\partial q_x}\right\}. \qquad (B56)$$

**Modified expression of $\sigma_{xx}(q_x,\omega)$**

By considering the symmetry of the summation indices $m$ and $n$, $\sigma_{xx}(q_x,\omega)$ can be further modified as

$$\sigma_{xx}(q_x,\omega)$$
$$=\frac{i}{\omega}\frac{e^2n_s}{m^*}\left[1+\frac{m^*}{2\pi\hbar^2 l_B^2 n_s q_x^2}\sum_{n,m}\frac{(n-m)\omega_c}{\omega+(n-m)\omega_c}(f_n-f_m)(n-m)(2\hbar\omega_c)F_{nm}(q_x)F_{mn}(-q_x)\right]$$
$$=\frac{i}{\omega}\frac{e^2n_s}{m^*}\left\{1+\frac{m^*}{2\pi\hbar^2 l_B^2 n_s q_x^2}\left[\sum_m f_m\sum_n\frac{(m-n)\omega_c}{\omega+(m-n)\omega_c}(m-n)(2\hbar\omega_c)F_{mn}(q_x)F_{nm}(-q_x)\right.\right.$$
$$\left.\left.-\sum_m f_m\sum_n\frac{(n-m)\omega_c}{\omega+(n-m)\omega_c}(n-m)(2\hbar\omega_c)F_{nm}(q_x)F_{mn}(-q_x)\right]\right\}. \qquad (B57)$$

Since the form factors can be expressed by using the harmonic-oscillator wavefunction $\varphi_m$ as

$$F_{mn}(q_x)=\langle\varphi_m|e^{-iq_xl_Bu}|\varphi_n\rangle, \qquad (B58)$$
$$F_{nm}(-q_x)=\langle\varphi_n|e^{iq_xl_Bu}|\varphi_m\rangle$$
$$=[F_{mn}(q_x)]^*, \qquad (B59)$$

and

$$F_{mn}(q_x)=F_{nm}(q_x), \qquad (B60)$$

we obtain

$$\sigma_{xx}(q_x,\omega)$$
$$=\frac{i}{\omega}\frac{e^2n_s}{m^*}\left[1+\frac{m^*}{\pi\hbar^2 l_B^2 n_s q_x^2}\sum_m f_m\sum_n\frac{(n-m)^2\omega_c^2}{\omega^2-(n-m)^2\omega_c^2}(n-m)(2\hbar\omega_c)\left|\langle\varphi_m|e^{-iq_xl_Bu}|\varphi_n\rangle\right|^2\right]. (B61)$$

Here, we consider

$$S_x(q_x,m)=2\sum_{n=0}^{\infty}(n-m)\hbar\omega_c\left|\langle\varphi_m|e^{-iq_xl_Bu}|\varphi_n\rangle\right|^2. \qquad (B62)$$

According to Bethe sum rule,

$$\left\langle \varphi_m \right| \left[ e^{\mp i q_x l_B u}, \left[ h_0, e^{\pm i q_x l_B u} \right] \right] \left| \varphi_m \right\rangle$$
$$= \sum_n (E_n - E_m) \left[ \left| \langle \varphi_m | e^{+i q_x l_B u} | \varphi_n \rangle \right|^2 + \left| \langle \varphi_m | e^{-i q_x l_B u} | \varphi_n \rangle \right|^2 \right]$$
$$= 2 \sum_n (n - m) \hbar \omega_c \left| \langle \varphi_m | e^{+i q_x l_B u} | \varphi_n \rangle \right|^2$$
$$= 2 \sum_n (n - m) \hbar \omega_c \left| \langle \varphi_m | e^{-i q_x l_B u} | \varphi_n \rangle \right|^2 . \qquad (B63)$$

Therefore,

$$S_x(q_x, m) = \frac{(\hbar q_x)^2}{m^*}. \qquad (B64)$$

Using this result, we obtain

$$\frac{m^*}{\pi \hbar^2 l_B^2 n_s q_x^2} \sum_m f_m \sum_{n=0}^{\infty} (n - m)(2\hbar\omega_c) \left| \langle \varphi_m | e^{-i q_x l_B u} | \varphi_n \rangle \right|^2 = \frac{1}{\pi l_B^2 n_s} \sum_m f_m$$
$$= 1. \qquad (B65)$$

Therefore,

$\sigma_{xx}(q_x, \omega)$

$$= i \frac{e^2 n_s}{m^*} \left[ \frac{m^*}{\pi \hbar^2 l_B^2 n_s q_x^2} \sum_m f_m \sum_n \frac{\omega}{\omega^2 - (n-m)^2 \omega_c^2} (n - m)(2\hbar\omega_c) F_{mn}(q_x) F_{nm}(-q_x) \right].$$
$$(B66)$$

By defining $l = n - m$, since $m \geq 0$ and $n \geq 0$,

$\sigma_{xx}(q_x, \omega)$

$$= i \frac{e^2 n_s}{m^*} \left[ \frac{2m^* \omega_c}{\pi \hbar l_B^2 n_s} \sum_{m=0}^{\infty} \sum_{l=-m}^{\infty} \left( \frac{l}{q_x^2} \right) \frac{\omega}{\omega^2 - (l\omega_c)^2} f_m F_{mn}(q_x) F_{nm}(-q_x) \right]. \qquad (B67)$$

By combining the positive and negative values of $l$, $\sigma_{xx}(q_x, \omega)$ can be rewritten as

$$\sigma_{xx}(q_x, \omega) = i \frac{e^2 n_s}{m^*} \sum_{l=1}^{\infty} \mathcal{F}(q_x, l) \left[ \frac{\omega}{\omega^2 - (l\omega_c)^2} \right], \qquad (B68)$$

where

$$\mathcal{F}_x(q_x, l) = \frac{2m^* \omega_c}{\pi \hbar l_B^2 n_s} \left( \frac{l}{q_x^2} \right) \sum_{m=0}^{\infty} (f_m - f_{m+l}) F_{m,m+l}(q_x) F_{m+l,m}(-q_x). \qquad (B69)$$

**Sum rule of $\mathcal{F}_x(q_x, l)$**

Next, we consider the sum rule of $\mathcal{F}_x(q_x, l)$.

$$\sum_{l=0}^{\infty} \mathcal{F}_x(q_x, l) = \frac{2m^* \omega_c}{\pi \hbar l_B^2 n_s q_x^2} \sum_{l=0}^{\infty} \sum_{m=0}^{\infty} l(f_m - f_{m+l}) F_{m,m+l}(q_x) F_{m+l,m}(-q_x) \qquad (B70)$$

The sum in Eq. (B70) can be rewritten as

$$\begin{aligned} &\sum_{l=0}^{\infty} \sum_{m=0}^{\infty} l(f_m - f_{m+l}) F_{m,m+l}(q_x) F_{m+l,m}(-q_x) \\ &= \sum_{l=0}^{\infty} \sum_{m=0}^{\infty} l(f_m - f_{m+l}) \left|F_{m,m+l}(q_x)\right|^2 \\ &= \sum_{m=0}^{\infty} \sum_{n=m}^{\infty} (n-m) f_m \left|F_{m,n}(q_x)\right|^2 - \sum_{m=0}^{\infty} \sum_{n=m}^{\infty} (n-m) f_n \left|F_{m,n}(q_x)\right|^2, \end{aligned} \qquad (B71)$$

where we introduced $n = m + l$. The second term on the right-hand side of Eq. (B71) can be rewritten by interchanging the summation indices.

$$\begin{aligned} \sum_{m=0}^{\infty} \sum_{n=m}^{\infty} (n-m) f_n \left|F_{m,n}(q_x)\right|^2 &= \sum_{n=0}^{\infty} \sum_{m=n}^{\infty} (m-n) f_m \left|F_{n,m}(q_x)\right|^2 \\ &= \sum_{n=0}^{\infty} \sum_{m=n}^{\infty} (m-n) f_m \left|F_{m,n}(q_x)\right|^2 \\ &= \sum_{m=0}^{\infty} \sum_{n=0}^{m} (m-n) f_m \left|F_{m,n}(q_x)\right|^2. \end{aligned} \qquad (B72)$$

Combining Eqs. (B71) and (B72), we obtain

$$\begin{aligned} &\sum_{n=m}^{\infty} \sum_{m=0}^{\infty} (n-m) f_m \left|F_{m,n}(q_x)\right|^2 - \sum_{n=m}^{\infty} \sum_{m=0}^{\infty} (n-m) f_n \left|F_{m,n}(q_x)\right|^2 \\ &= \sum_{n=m}^{\infty} \sum_{m=0}^{\infty} (n-m) f_m \left|F_{m,n}(q_x)\right|^2 + \sum_{m=0}^{\infty} \sum_{n=0}^{m} (n-m) f_m \left|F_{m,n}(q_x)\right|^2 \\ &= \sum_{m=0}^{\infty} f_m \sum_{n=0}^{\infty} (n-m) \left|F_{m,n}(q_x)\right|^2 \\ &= \sum_{m=0}^{\infty} f_m \sum_{n=0}^{\infty} (n-m) \left|\langle \varphi_m | e^{-iq_x l_B u} | \varphi_n \rangle\right|^2. \end{aligned} \qquad (B73)$$

Using Eq. (B65) and substituting Eq. (B73) into Eq. (B70), we obtain

$$\sum_{l=0}^{\infty} \mathcal{F}_x(q_x, l) = 1. \qquad (B74)$$

## Appendix C: Dispersion relation from the Hopfield Hamiltonian

We derive the dispersion relation associated with the Hopfield Hamiltonian used in the main text. Following Ref. [28], the Hamiltonian is

$$\hat{\mathcal{H}}/\hbar = \omega_{\mathrm{MP}}\hat{a}^\dagger\hat{a} + \sum_{n=2}\left[n\omega_c\hat{b}_n{}^\dagger\hat{b}_n + ig_n(\hat{a}+\hat{a}^\dagger)\left(\hat{b}_n - \hat{b}_n{}^\dagger\right)\right] + D(\hat{a}+\hat{a}^\dagger)^2. \qquad (C1)$$

We define the quadrature operators

$$\hat{Q} = \hat{a} + \hat{a}^\dagger, \qquad (C2)$$

$$\hat{Y}_n = i\left(\hat{b}_n - \hat{b}_n{}^\dagger\right), \qquad (C3)$$

$$\hat{\Pi} = i(\hat{a} - \hat{a}^\dagger), \qquad (C4)$$

$$\hat{X}_n = \hat{b}_n + \hat{b}_n{}^\dagger. \qquad (C5)$$

Using the quadrature operators, the Hamiltonian can be rewritten as

$$\hat{\mathcal{H}}/\hbar = \hbar\omega_{\mathrm{MP}}\hat{a}^\dagger\hat{a} + \sum_{n=2}\left(n\hbar\omega_c\hat{b}_n{}^\dagger\hat{b}_n + g_n\hat{Q}\hat{Y}_n\right) + D\hat{Q}^2. \qquad (C6)$$

The Heisenberg equations of motion for $\hat{Q}$ and $\hat{Y}_n$ are given by

$$\frac{d}{dt}\hat{Q} = -\omega_{\mathrm{MP}}\hat{\Pi}, \qquad (C7)$$

$$\frac{d}{dt}\hat{Y}_n = -n\omega_c\hat{X}_n. \qquad (C8)$$

Similarly, the Heisenberg equations for $\hat{\Pi}$ and $\hat{X}_n$ are given by

$$\frac{d}{dt}\hat{\Pi} = (\omega_{MP} + 4D) + 2\sum_{n=2} g_n\hat{Y}_n, \qquad (C9)$$

$$\frac{d}{dt}\hat{X}_n = -n\omega_c\hat{Y}_n - 2g_n\hat{Q}. \qquad (C10)$$

Eliminating $\hat{\Pi}$, $\hat{X}_n$, and $\hat{Y}_n$ in frequency space gives

$$\left[\omega^2 - \omega_{MP}(\omega_{\mathrm{MP}} + 4D) - \sum_{n=2}\frac{2g_n^2\omega_{\mathrm{MP}}(n\omega_c)}{\omega^2 - (n\omega_c)^2}\right]\hat{Q} = 0. \qquad (C11)$$

Thus, the Hopfield model gives the dispersion relation

$$\frac{\omega_{MP}^2}{\omega^2} = 1 - \frac{1}{\omega^2}\left[4\omega_{\mathrm{MP}}D + \sum_{n=2}\frac{4g_n^2\omega_{\mathrm{MP}}(n\omega_c)}{\omega^2 - (n\omega_c)^2}\right]. \qquad (C12)$$

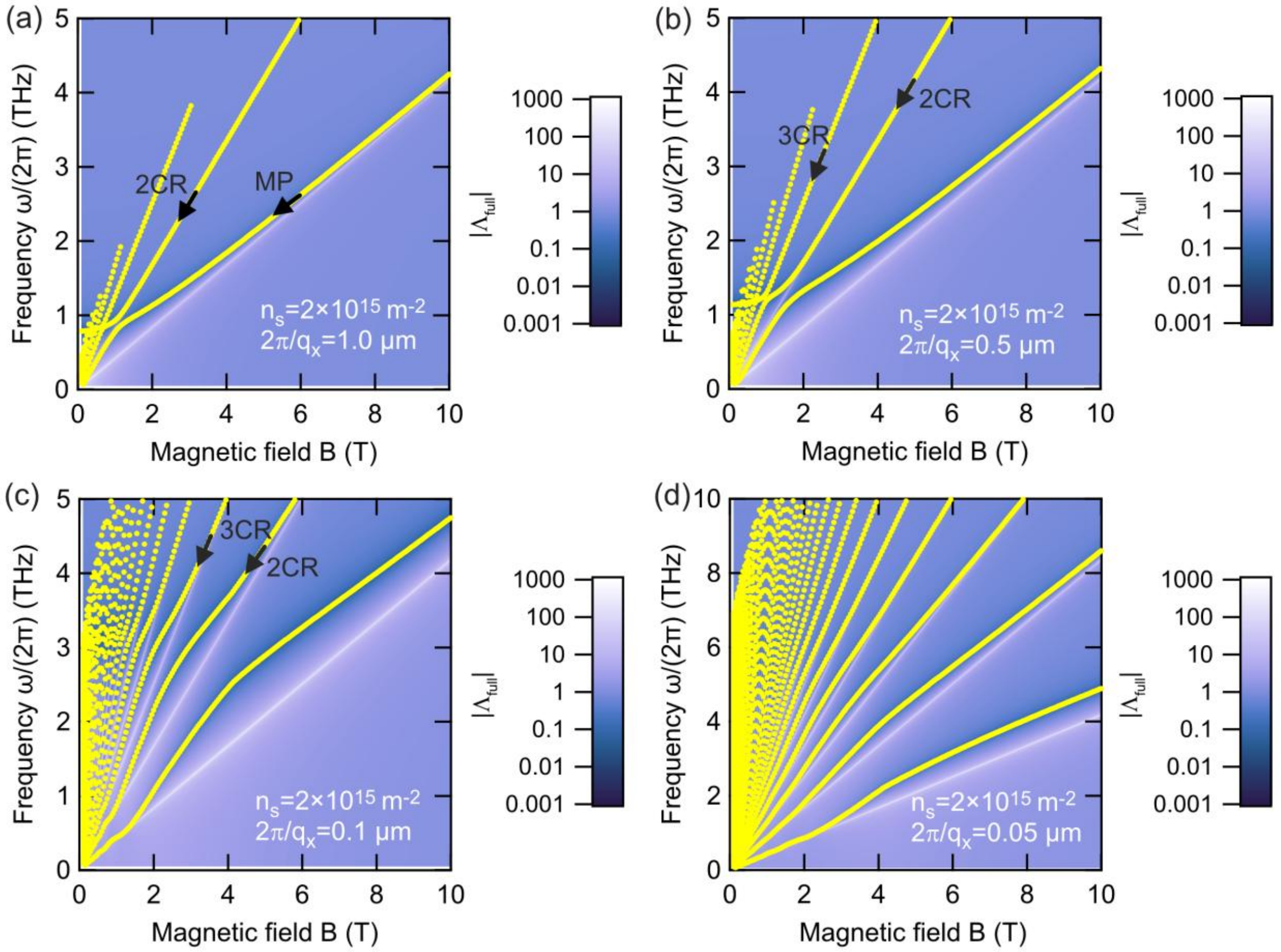

Fig. 5. Comparison between the full and approximate dispersions. The yellow circles are the solutions of $|\Lambda_{\mathrm{approx}}(q_x, \omega)| = 0$, overlaid on the same color-coded maps of $|\Lambda_{\mathrm{full}}(q_x, \omega)|$ as in Fig. 2.

## Appendix D: Approximate dispersion relation in Eq. (21)

To assess the accuracy of the approximate dispersion relation in Eq. (21), we overlay its solutions on the color maps in Fig. 2 (see the yellow circles in Fig. 5). The resulting dispersion curves closely reproduce those obtained from the full dispersion relation in Eq. (16), confirming the validity of the approximation.